# Algebraic Operators for Querying Pattern Bases

Rokia Missaoui[1], Léonard Kwuida[1], Mohamed Quafafou[2], Jean Vaillancourt[1]

[1] Université du Québec en Outaouais
Gatineau (Québec) Canada, J8X 3X7
[2] LSIS - CNRS UMR 6168
Université Aix-Marseille, France
{rokia.missaoui,leonard.kwuida,jean.vaillancourt}@uqo.ca[*],
mohamed.quafafou@univmed.fr

**Abstract.** The objectives of this research work which is intimately related to pattern discovery and management are threefold: (i) handle the problem of pattern manipulation by defining operations on patterns, (ii) study the problem of enriching and updating a pattern set (e.g., concepts, rules) when changes occur in the user's needs and the input data (e.g., object/attribute insertion or elimination, taxonomy utilization), and (iii) approximate a "presumed" concept using a related pattern space so that patterns can augment data with knowledge. To conduct our work, we use formal concept analysis (FCA) as a framework for pattern discovery and management and we take a joint database-FCA perspective by defining operators similar in spirit to relational algebra operators, investigating approximation in concept lattices and exploiting existing work related to operations on contexts and lattices to formalize such operators.

## 1 Introduction

The recent research topic of pattern discovery and management refers to a set of activities related to the extraction, description, manipulation and storage of patterns in a similar (but more elaborated) way as data are managed by database applications. In pattern management and inductive databases [4,5,16,24], patterns are knowledge artifacts (e.g., association rules, clusters) extracted from data using data mining procedures (generally run in advance), and retrieved upon user's request. A pattern is then a concise and semantically rich representation of raw data. An example of a pattern could be a cluster that represents a set of Star Alliance members with their common features (e.g., fleet size, set of destinations).

In many database and data warehouse applications, users tend to be drowning in data and even in patterns while they are actually interested in a very limited set of knowledge pieces. Moreover, the scope of patterns to explore differs from one user to another and changes over time. Finally, one is frequently interested in an exploratory and iterative process of data mining (DM) to discover patterns under different scenarios and different hypotheses. In order to reduce the memory overload of the user and his working space induced by the large set of mined patterns, we propose to define a set of algebraic operators similar in spirit to operators of relational algebra. Such operators will allow "data mining on demand" (i.e., data mining according to user's needs and perspectives) and rely on key operations on concept lattices such as selection, projection and join. Additional operations will be defined either to enrich the pattern basis or to identify the patterns that best approximate a "presumed" concept.

[*] partially supported by the Natural Sciences and Engineering Research Council of Canada (NSERC).

The following example is an elementary way to display information. It presents the Star Alliance members in year 2000 with their destinations [8].

| Star Alliance | Latin America | Europe | Canada | Asia Pacific | Middle East | Africa | Mexico | Caribbean | US |
|---|---|---|---|---|---|---|---|---|---|
| Air Canada | × | × | × | × | × |  | × | × | × |
| Air New Zealand |  | × |  | × |  |  |  |  | × |
| All Nippon Airways |  | × |  | × |  |  |  |  | × |
| Ansett Australia |  |  |  | × |  |  |  |  |  |
| The Austrian Airlines Group |  | × | × | × | × | × |  |  | × |
| British Midland |  | × |  |  |  |  |  |  |  |
| Lufthansa | × | × | × | × | × | × | × |  | × |
| Mexicana | × |  | × |  |  |  | × | × | × |
| Scandinavian Airlines | × | × |  | × |  | × |  |  | × |
| Singapore Airlines |  | × | × | × | × | × |  |  | × |
| Thai Airways International | × | × |  | × |  |  |  | × | × |
| United Airlines | × | × | × | × |  |  | × | × | × |
| VARIG | × | × |  | × |  | × | × |  | × |

**Fig. 1.** Star Alliance members and their flying destinations in year 2000.

## 2  Contexts, Concept Lattices and their Ideals

Formal Concept Analysis is a branch of applied mathematics, which is based on a formalization of concept and concept hierarchy [9]. It has been successfully used for conceptual clustering and rule generation. Let $\mathbb{K} = (G, M, I)$ be a formal context, where $G$, $M$ and $I$ stand for a set of objects, a set of attributes, and a binary relation between $G$ and $M$ respectively. Two functions, $f_1$ and $f_2$, summarize the links between subsets of objects and subsets of attributes induced by $I$. Function $f_1$ maps a set of objects into a set of common attributes, whereas $f_2$ is the dual for attribute sets:

$$f_1 : \mathcal{P}(G) \to \mathcal{P}(M), A \mapsto A' := \{a \in M \mid \forall o \in A, oIa\},$$
$$f_2 : \mathcal{P}(M) \to \mathcal{P}(G), B \mapsto B' := \{o \in G \mid \forall a \in B, oIa\}.$$

Furthermore, the compound operators $f_2 \circ f_1$ and $f_1 \circ f_2$ (denoted by $''$) are *closure* operators on $G$ and $M$ respectively. In particular, $Z \subseteq Z''$ and $(Z'')'' = Z''$ for $Z \in \mathcal{P}(M) \cup \mathcal{P}(G)$. The set $Z$ is closed if $Z'' = Z$. A formal concept $c$ is a pair of sets $(A, B)$ with $A \subseteq G, B \subseteq M$, $A = B'$ and $B = A'$. $A$ is called the extent of $c$ (denoted by $\text{ext}(c)$) and $B$ its intent (denoted by $\text{int}(c)$). In the closed *itemset* mining framework [18,29], $A$ and $B$ correspond to the notion of closed *tidset* and closed *itemset* respectively. The set of all concepts of $\mathbb{K}$ is denoted by $\mathfrak{B}(\mathbb{K})$. Ordered by $(A, B) \leq (C, D) : \iff A \subseteq C$, it forms a complete lattice[3] called the concept lattice of $\mathbb{K}$ and denoted by $\underline{\mathfrak{B}}(\mathbb{K})$. For $(A, B)$ and $(C, D)$ in $\underline{\mathfrak{B}}(\mathbb{K})$ we have

$$(A, B) \vee (C, D) = ((A \cup C)'', B \cap D) \text{ and } (A, B) \wedge (C, D) = (A \cap C, (B \cup D)'').$$

The sets $G$ and $M$ are related to $\underline{\mathfrak{B}}(G, M, I)$ by the following mappings, where $x'$ stands for $\{x\}'$ with $x \in G \cup M$.

$$\gamma : G \to \underline{\mathfrak{B}}(G, M, I) \quad \text{and} \quad \mu : M \to \underline{\mathfrak{B}}(G, M, I)$$
$$g \mapsto \gamma g := (g'', g') \quad \quad m \mapsto \mu m := (m', m''),$$

with $gIm \iff \gamma g \leq \mu m$. The $\gamma g$'s and the $\mu m$'s form the building blocks of the concept lattice. In fact, any concept is the join of some $\gamma g$'s and the meet of some $\mu m$'s; i.e. if $c$ is a concept of $(G, M, I)$, then there are sets $C_1 \subseteq G$ and $C_2 \subseteq M$ such that $c = \bigvee\{\gamma g \mid g \in C_1\}$ and $c = \bigwedge\{\mu m \mid m \in C_2\}$. We call $C_1$ a generator of the extent of $c$, and $C_2$ a generator of the intent of $c$. In fact $C_1'' = \text{ext}(c) = \{g \in G \mid \gamma g \leq c\}$ and $C_2'' = \text{int}(c) = \{m \in M \mid \mu m \geq c\}$.

Ideals and filters play an important role in describing *selection* and *approximation* on concepts. An *order ideal* is a downward closed subposet. For a poset $(P, \leq)$ and $X \subseteq P$, the intersection of all order ideals containing $X$ is the smallest order ideal containing $X$. It is called the *order ideal generated by $X$* and denoted by $\downarrow X$. If $X = \{a\}$ then $\downarrow a := \downarrow X = \{x \in P \mid x \leq a\}$, and is called *principal ideal*. Dually, an *order filter* is an upward closed subposet. For $X \subseteq P$, the intersection of all order filters containing $X$ is an order filter containing $X$, called the *order filter generated by* $X$. If $X = \{a\}$ then $\uparrow a := \uparrow X = \{x \in P \mid x \geq a\}$ is called principal filter. In a lattice, a *lattice ideal* or simply *ideal* is an order ideal closed under finite suprema. For $X \subseteq P$, the intersection of all lattice ideals containing $X$ is again a lattice ideal containing $X$, called the *ideal generated by* $X$. It always contains the order ideal generated by $X$. All principal ideals are also lattice ideals. If $x, y$ are incomparable in $L$, then the order ideal generated by $\{x, y\}$ is $\downarrow x \cup \downarrow y$ which is not a lattice ideal. It is smaller than the lattice ideal generated by $\{x, y\}$. The notion of *lattice filter* or simply *filter* is defined dually [7].

---

[3] This is a poset in which every subset $X$ has an infimum ($\bigwedge X$) and a supremum ($\bigvee X$). We set $a \wedge b := \bigwedge\{a, b\}$ and $a \vee b := \bigvee\{a, b\}$.

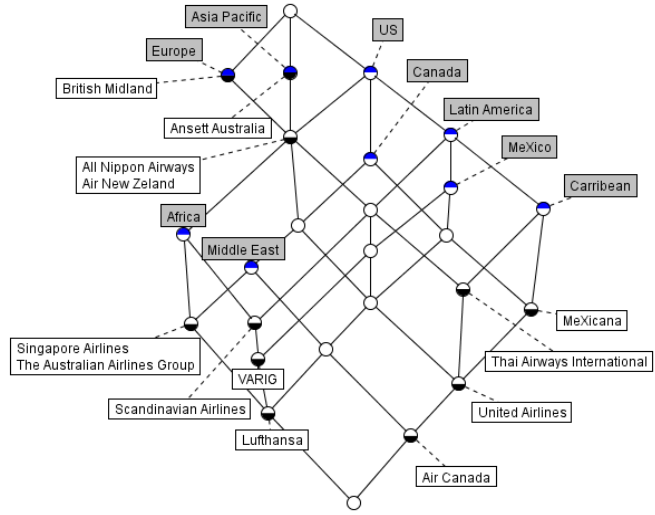

**Fig. 2.** Concept lattice of the context of Figure 1.

## 3 Information Systems

Frequently, data are not directly encoded in a "binary" form, but rather as a many-valued context in the form of a tuple $(G, M, W, I)$ of sets such that $I \subseteq G \times M \times W$, with $(g, m, w_1) \in I$ and $(g, m, w_2) \in I$ imply $w_1 = w_2$. $G$ is called the set of objects, $M$ the set of attributes (or attribute names) and $W$ the set of attribute values. If $(g, m, w) \in I$, then $w$ is the value of the attribute $m$ for the object $g$. Another notation is $m(g) = w$ where $m$ is a partial map from $G$ to $W$. Many-valued contexts can be transformed into binary contexts, via conceptual scaling. A *conceptual scale* for an attribute $m$ of $(G, M, W, I)$ is a binary con-text $\mathbb{S}_m := (G_m, M_m, I_m)$ such that $m(G) \subseteq G_m$. Intuitively, $M_m$ discretizes or groups the attribute values into $m(G)$, and $I_m$ describes how each attribute value $m(g)$ is related to the elements in $M_m$. For an attribute $m$ of $(G, M, W, I)$ and a conceptual scale $\mathbb{S}_m$ we derive a binary context $\mathbb{K}_m := (G, M_m, I^m)$ with $gI^m s_m :\iff m(g)I_m s_m$, where $s_m \in M_m$. This means that an object $g \in G$ is in relation with a scaled attribute $s_m$ iff the value of $m$ on $g$ is in relation with $s_m$ in $\mathbb{S}_m$. With a conceptual scale for each attribute we get the *derived context* $\mathbb{K}^S := (G, N, I^S)$ where $N := \bigcup \{M_m \mid m \in M\}$ and $gI^S s_m \iff m(g)I^m s_m$. In practice, the set of objects remains unchanged; each attribute name $m$ is replaced by the scaled attributes $s_m \in M_m$. An *information system* is a many-valued context $(G, M, W, I)$ with a set of scales $(\mathbb{S}_m)_{m \in M}$. The choice of a suitable set of scales depends on the interpretation, and is usually done with the help of a domain expert. A *Conceptual Information System* is a one-valued (or many-valued) context together with a set of conceptual scales (or hierarchies). Such a set of scales is called *conceptual schema* [21,22]. Other scaling methods have also been proposed (see for e.g., [19,20]).

## 4 Relations and Relation Schema

We first recall key notions on relation schema and relational algebra [15,1]. A relation scheme is a set of attribute names $R := \{A_1, \ldots, A_n\}$, denoted by $R(A_1, \ldots, A_n)$ or simply $R$, where $n \in \mathbb{N}$ is the arity of $R$. For each attribute name $A_i$, there is a set $\text{dom}A_i$, called the domain of $A_i$. A relation on the scheme $R$, denoted by $r(R)$ or $r(A_1, \ldots, A_n)$ or simply $r$, is a set[4] $\{t_1, \ldots, t_p\}$ of mappings from $R$ to $D := \bigcup \{\text{dom}A_i \mid 1 \leq i \leq n\}$ such that $t(A_i) \in \text{dom}A_i$ for any tuple $t \in r$. The $A$-value of $t$ is $t(A)$, and more generally, the Y-value of $t$ is t(Y), where $Y \subseteq \{A_1, \ldots, A_n\}$. A relation $r$ can be interpreted as a table, where the rows are its tuples and the columns are headed by the attribute names. We assume that each relation $r$ has an attribute name $K$, that is a key (i.e., for $i, j \in \{1, \ldots, p\}$, $t_i \neq t_j$ iff $t_i(K) \neq t_j(K)$). Then a relation $r$ on a scheme $R$ is nothing else than a many-valued context $(G, R, D, I)$ with

$$G := \{t(K) \mid t \in r\} \quad \text{and} \quad (t(K), m, w) \in I :\iff t(m) = w.$$

Each many-valued context can be transformed into a binary context using a set of scales $(\mathbb{S}_m)_{m \in M}$ [9]. Therefore, to each relation $r$ we can associate a formal context $\mathbb{K}(r)$ and a concept lattice $\mathfrak{B}(r)$, which depends on the chosen scales $(\mathbb{S}_m)_{m \in M}$.

To handle the information stored in relations, a relational algebra has been defined to allow operations on relations (see for example [1,15]). Our aim is to see how these operations can be encoded on concept lattices. Before we proceed, let us first recall some key operations. We start with a logic on the attributes. Let $r$ be a relation on a schema $R$. The atomic formulae are of the form $A \dot{=} a$ for $A \in R$ and $a \in \text{dom} A$. The connectors $\wedge$, $\vee$ and $\neg$ are defined as usual. Examples of formulae are

$$\varphi_1 := (A \dot{=} a \wedge B \dot{=} b); \quad \varphi_2 := (A \dot{=} a \vee B \dot{=} b) \quad \text{and} \quad \varphi_3 := \neg(A \dot{=} a).$$

---
[4] All relations considered here are finite.

**Selection.** We consider a relation $r(A_1, \ldots, A_n)$ and $(G, M, I)$ a binary context derived from $r$ via the scales $(\mathbb{S}_\mathbb{A})_{A \in R}$. We are interested in a relation whose tuples are those of $r$ with a certain value $a$ on a specified attribute $A_j$. This is the *selection* operation denoted by $\text{Select}(r, A_j \dot{=} a)$ or $\sigma_{A_j \dot{=} a}(r)$ and defined by

$$\text{Select}(r, A_j \dot{=} a) := \{t \in r \mid t(A_j) = a\}.$$

This operation gives a special sub-context of $(G, M, I)$ with all attributes from $M$ and all those objects in $G$ that have the scaled attribute[5] $a$. On the concept lattice side, the above selection operation can be expressed as $\text{Select}(\mathfrak{B}(r), A_j \dot{=} a)$ and corresponds to the order ideal $\downarrow \mu a$. In case of a conjunctive condition on atomic formulae, $\varphi_\wedge := \bigwedge \{A_j \dot{=} a_j \mid j \in J \subseteq \{1, \ldots, n\}\}$, the operation $\text{Select}(\mathfrak{B}(r), \varphi_\wedge)$ gives the ideal $\downarrow \bigwedge_{j \in J} \mu a_j$. For a disjunctive condition on atomic formulae, $\varphi_\vee := \bigvee \{A_j \dot{=} a_j \mid j \in J \subseteq \{1, \ldots, n\}\}$, the operation $\text{Select}(\mathfrak{B}(r), \varphi_\vee)$ gives the order ideal $\bigcup_{j \in J} \downarrow \mu a_j$. The case of negation is to be handled with good care. For a concept $(A, B)$ of $(G, M, I)$, its *weak negation* (resp. *weak opposition*) is the concept $((G \setminus A)'', (G \setminus A)')$ (resp. $((M \setminus B)', (M \setminus B)'')$ [13,27,28]. $\text{Select}(\mathfrak{B}(r), \neg(A \dot{=} a))$ will produce all concepts of $\mathbb{K}(r)$ whose objects do not have the $A$-value $a$. This is probably not an order ideal of $\mathfrak{B}(r)$. If $G \setminus \{t(K) \mid t(A) = a\}$ is closed, then $\text{Select}(\mathfrak{B}(r), \neg(A \dot{=} a))$ is an order ideal. Otherwise, the output of $\text{Select}(\mathfrak{B}(r), \neg(A \dot{=} a))$ is only a sub-hierarchy of the order ideal generated by $\gamma(G \setminus \{t(K) \mid t(A) = a\})$.

**Projection.** Let $r(A_1, \ldots, A_n)$ be a relation and $Y \subseteq \{A_1, \ldots, A_n\}$. Then $\text{Project}(r, Y) = \Pi_Y(r) = \{t(Y) \mid t \in r\}$ restricts the tuples of $r$ to the attributes in $Y$. The projection defined above is equivalent to having a sub-context $(G, N, J)$ of $(G, M, I)$ where $N$ is the set of scaled values of the attributes in $Y$. Two concepts $c_1$ and $c_2$ in $\mathfrak{B}(r)$ are *Y-equivalent* if $\text{int}(c_1) \cap Y = \text{int}(c_2) \cap Y$.

---

[5] If $\mathbb{S}_{A_j}$ is not a nominal scale, some objects that do not have exactly the $A_j$-value $a$ can also be chosen, provided their $A_j$-value is in relation with the scaled attribute $s_a$, that represents the group in which $a$ belongs.

Every concept $c$ has a greatest concept equivalent to it. Then Project$(r, Y)$ translates in FCA into a projection on concepts Project$(\mathfrak{B}(r), Y)$ which is a copy of the sub-hierarchy

$$R_Y(r) := (\{c \in \mathfrak{B}(r) \mid c \text{ is the greatest element of its } Y\text{-equivalence class}\}, \leq),$$

and is a $\wedge$-subsemilattice of $\mathfrak{B}(r)$. More precisely

$$\text{Project}(\mathfrak{B}(r), Y) = \{((int(c) \cap Y)', int(c) \cap Y) \mid c \in R_Y(r)\} = \underline{\mathfrak{B}}(G, N, J).$$

Figure 3 illustrates a projection on $Y := \{$Canada, Asia pacific$\}$ and a selection. The four $Y$-equivalence classes induced by this projection are identified (left) and their order displayed (right). The structure inside the lowest class (left) displays the selection of Star Alliance members whose destinations include Canada and Asia Pacific.

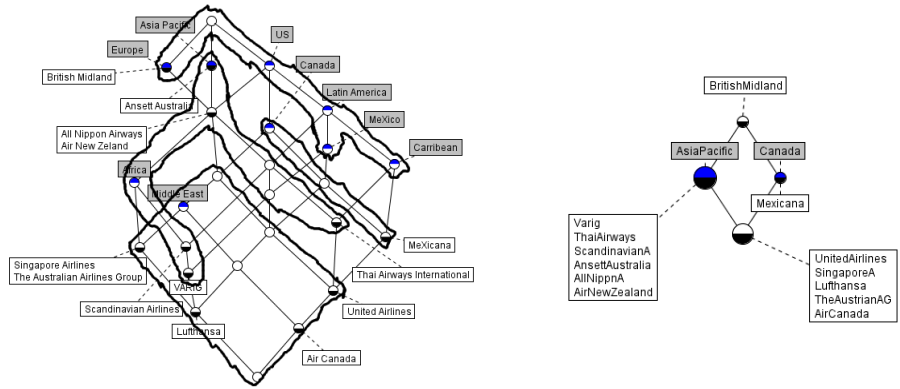

**Fig. 3.** Projection and selection.

**Natural Join.** The most frequently used binary operation is the natural join. Let $r(R)$ and $s(S)$ be two relations. The natural join of $r$ and $s$, denoted by $r \bowtie s$ or Join$(r, s)$, is the relation $q(T)$ with schema $T = R \cup (S \setminus R)$ containing all tuples $t$ over $T$ such that there are tuples $t_1 \in r$ and $t_2 \in s$ satisfying $t(R) = t_1$ and $t(S) = t_2$.

$$\text{Join}(r, s) := \{t_1 | t_2 \text{ for } t_1 \in r \text{ and } t_2 \in s \text{ such that } t_1(R \cap S) = t_2(R \cap S)\},$$

where $t_1|t_2$ denotes the tuple of a relation on the scheme $R \cup (S \setminus R)$ defined for all attribute name $A$ by

$$(t_1|t_2)(A) := \begin{cases} t_1(A) & \text{if } A \in R \\ t_2(A) & \text{if } A \in S \setminus R. \end{cases}$$

Note that $t_1|t_2$ is defined only for tuples that coincide on $R \cap S$ [6]. Those tuples are called *joinable*. If $R \cap S$ is empty (i.e., there are no common attributes between the two relation schemes), then $t_1|t_2$ is defined for all tuples of $r$ and of $s$, and $\text{Join}(r,s)$ gives the Cartesian product $r \times s$. When the join attribute is the identifier of objects in two contexts, then the relational join operation is equivalent to the apposition (see Subsection 5.2) of $\mathbb{K}_1 := (G, M_1, I_1)$ and $\mathbb{K}_2 := (G, M_2, I_2)$ having the same set of objects to get a context $\mathbb{K} := (G, M_1 \cup M_2, I_1 \cup I_2)$. The lattice corresponding to the context $\mathbb{K}$ is isomorphic to a subdirect product of the $\vee$-semilattices $\mathfrak{B}(\mathbb{K}_1)$ and $\mathfrak{B}(\mathbb{K}_2)$ and can be expressed by nested line diagrams [9].

Other binary operations on relations include the set-based operations such as the intersection, the union and the difference ($r \cap s, r \cup s, r \setminus s$). These operations will not be discussed in this paper. A query in the form of a relational algebra expression [15] is a well-formed expression that contains a finite number of relational algebra operators whose operands are relations.

## 5 Algebraic Operations on Contexts

In the following, we define a kernel of main operations on contexts and lattices.

### 5.1 Sub-contexts: local views of a context

Given a context $\mathbb{K}$, one can choose (or is required) to have a local view, by restricting it to some objects and some attributes of $\mathbb{K}$. In this subsection we show how the concepts of a sub-context are related to those of the initial context (see for example [9, Section 3.1]). Formally, for a context $\mathbb{K} := (G, M, I)$, a *sub-context* is a triple $\mathbb{H} := (H, N, J)$ such that $H \subseteq G$, $N \subseteq M$ and $J = I \cap (H \times N)$. The hierarchy of concepts in $\mathbb{H}$ can be seen as a sub-hierarchy of the concepts in $\mathbb{K}$. For each concept $\mathfrak{u} := (U, V)$ of $\mathbb{H}$, its extent $U$ is also a subset of $G$ and its closure $U''$ in $\mathbb{K}$ induces a concept $(U'', U')$. The intent $V$ is also a subset of $M$ and its closure $V''$ in $\mathbb{K}$ induces a concept $(V', V'')$. We set $\varphi_1 \mathfrak{u} := (U'', U')$ and $\varphi_2 \mathfrak{u} := (V', V'')$. Then $\varphi_1$ and $\varphi_2$ define two mappings from $\mathfrak{B}(\mathbb{H})$ to $\mathfrak{B}(\mathbb{K})$ that are order embeddings i.e. for $1 \leq i \leq 2$ we have

$$\forall \mathfrak{u}, \mathfrak{v} \in \mathfrak{B}(\mathbb{H}), \begin{cases} \mathfrak{u} \leq \mathfrak{v} \implies \varphi_i \mathfrak{u} \leq \varphi_i \mathfrak{v}, & \text{preserving the order of } \mathfrak{B}(\mathbb{H}) \\ \varphi_i \mathfrak{u} \leq \varphi_i \mathfrak{v} \implies \mathfrak{u} \leq \mathfrak{v}, & \text{reflecting the order of } \mathfrak{B}(\mathbb{K}). \end{cases}$$

---

[6] We assume that the join attributes have the same name in the two tables $R$ and $S$.

To avoid confusion, we usually replace the derivation $'$ by the name of the relation of the context in which the derivation is done. Thus, for a subcontext $(H, N, J)$ of $(G, M, I)$, $U \subseteq H$ and $V \subseteq N$ we have

$$U^J = U^I \cap N \text{ and } V^J = V^I \cap H. \qquad (\dagger)$$

Next we compare the closures in $\mathbb{H}$ and in $\mathbb{K}$. From the equalities in $(\dagger)$ above, we have $U^J \subseteq U^I$ and $U^{JI} \supseteq U^{II}$. Therefore

$$U^{JJ} = U^{JI} \cap H \supseteq U^{II} \cap H \supseteq U. \qquad (\ddagger)$$

This means that the closure of a set $U$ of objects in $\mathbb{H}$ is usually larger than the restriction on $H$ of its closure in $\mathbb{K}$. If $U$ is closed in $\mathbb{H}$, we have the equality, since $(\ddagger)$ leads to $U = U^{JJ} \supseteq U^{II} \cap H \supseteq U$ and $U = U^{II} \cap H$. Similarly we get $V^{II} \cap N \subseteq V^{JJ}$ and $V^{II} \cap N = V$ if $V$ is closed in $\mathbb{H}$. The context $\mathbb{K}$ in Figure 4 and the sub-context $\mathbb{H}$ given by $H := \{2, 3, 5, 6\}$ and $N := \{a, b, d, f\}$ illustrate the inclusion above. For $U := \{2, 3\}$ we have $U^I = \{a, b, d, e\}$ and $U^{II} = \{2, 3, 4\}$ as well as $U^J = \{a, b, d\}$ and $U^{JJ} = \{2, 3, 5\}$. Thus the closures in $\mathbb{H}$ and $\mathbb{K}$ are not always comparable, but only their restrictions on $H$. We have $U^{II} \cap H = \{2, 3\} \subseteq U^{JJ}$. In $\mathbb{H}$, $\{5\}$ is closed since $5^{JJ} = \{5\}$; also $5^{II} = \{4, 5\}$ and $5^{II} \cap H = \{5\} = 5^{JJ}$. We have seen that if $U$ is an extent and $V$ an intent of $\mathbb{H}$, then $U^{II} \cap H = U$ and $V^{II} \cap N = V$. Thus if $(U, V)$ is a concept of $\mathbb{H}$, then $(U^{II}, U^I)$ and $(V^I, V^{II})$ are concepts of $\mathbb{K}$ whose restrictions on $\mathbb{H}$ are equal to $(U, V)$; i.e. $(U^{II} \cap H, U^I \cap N) = (U, V)$ and $(V^I \cap H, V^{II} \cap N) = (U, V)$. Note that $U^{II} \subseteq V^I$ and $(U^{II}, U^I) \leq (V^I, V^{II})$. Therefore, for every concept $(A, B)$ of $\mathbb{K}$ in the interval $\left[(U^{II}, U^I); (V^I, V^{II})\right]$ we have $(A \cap H, B \cap N) = (U, V)$. This means that the concepts of $\mathbb{H}$ can be identified with some intervals of $\mathbb{K}$ (see Section 6). If every concept of $\mathbb{K}$ is in such an interval then the restriction on $\mathbb{H}$ of every concept $(A, B)$ of $\mathbb{K}$ is a concept $(A \cap H, B \cap N)$ of $\mathbb{H}$. The mapping

$$\Pi_{\mathbb{H}} : \mathfrak{B}(G, M, I) \to \mathfrak{B}(H, N, J)$$
$$(A, B) \mapsto (A \cap H, B \cap N)$$

is a surjective complete lattice homomorphism. The mapping $\Pi_{\mathbb{H}}$ from $\mathfrak{B}(\mathbb{K})$ to $\mathfrak{B}(\mathbb{H})$ or simply from the initial context $\mathbb{K}$ to its sub-context $\mathbb{H}$ coincides with a mixture of two relational operations: a selection of tuples $t$ with $t(K) \in H$ and a projection on attributes in $N$. Therefore, the mixture of selection and projection operations commonly used in relational databases can be implemented for concept lattice manipulation by means of the theory described above.

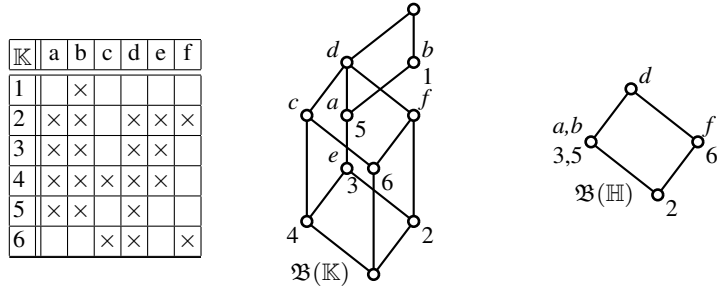

**Fig. 4.** A context $\mathbb{K}$, its concept lattice $\mathfrak{B}(\mathbb{K})$ and a concept lattice of its sub-context $\mathfrak{B}(\mathbb{H})$.

### 5.2 Enlarging contexts

The reverse situation of local views is enlarging the context. One possibility is to enlarge the set of attributes. Two contexts $\mathbb{K}_1 := (G, M_1, I_1)$ and $\mathbb{K}_2 := (G, M_2, I_2)$, with the same set of objects can be combined to get a context $\mathbb{K} := (G, M_1 \uplus M_2, I_1 \cup I_2)$. The context $\mathbb{K}$ is called the *apposition* of $\mathbb{K}_1$ and $\mathbb{K}_2$, and denoted by $\mathbb{K}_1 | \mathbb{K}_2$. The extent of the concepts in the resulting lattice is exactly the intersection of extents of $\mathbb{K}_1$ and $\mathbb{K}_2$ [26]. In general two contexts $(G_1, M_1, I_1)$ and $(G_2, M_2, I_2)$ can be put together to get a context $(G, M, I)$ with $G := G_1 \cup G_2$, $M := M_1 \cup M_2$ and $I := I_1 \cup I_2$ if they agree on their intersections i.e. for any $g \in G_1 \cap G_2$ and $m \in M_1 \cap M_2$ we have $gI_1m \iff gI_2m$. Here we have $g \in G_1 \setminus G_2$ and $m \in M_2 \setminus M_1$ imply $(g, m) \notin I$ as well as $g \in G_2 \setminus G_1$ and $m \in M_1 \setminus M_2$ imply $(g, m) \notin I$. Note that $(G_1, M_1, I_1) \subseteq (G, M, I) \supseteq (G_2, M_2, I_2)$.

Another possibility is to enlarge the set of objects. Two contexts $\mathbb{K}_1 := (G_1, M, I_1)$ and $\mathbb{K}_2 := (G_2, M, I_2)$ (having the same set of attributes) can be combined to get a context $\mathbb{K} := (G_1 \uplus G_2, M, I_1 \cup I_2)$. The context $\mathbb{K}$ is called the *subposition* of $\mathbb{K}_1$ and $\mathbb{K}_2$, and denoted by $\frac{\mathbb{K}_1}{\mathbb{K}_2}$. Such operation is useful to incrementally update the lattice $\mathfrak{B}(\mathbb{K}_1)$ when a set $G_2$ of objects is added [25].

### 5.3 Generalized Patterns

The objective here is to exploit generalization hierarchies attached to properties to get a lattice with more abstract concepts. Producing generalized patterns from concept lattices when a taxonomy on attributes is provided can be done in different ways with distinct performance costs that depend on the peculiarities of the input (e.g., size, density) and the operations used. One way consists to use context apposition to conduct the assembly (join) of the initial lattice of non generalized attributes (e.g., destinations of airline companies) with the lattice corresponding to the taxonomy of attributes (e.g., city, country, continent), and then perform a projection of the resulting lattice on the generalized attributes (e.g., country) only. Figure 5 shows generalized patterns when the attributes *Canada* and *US* are replaced with the generalized attribute *North America*, and *Mexico* and *Latin America* are replaced with *South America*.

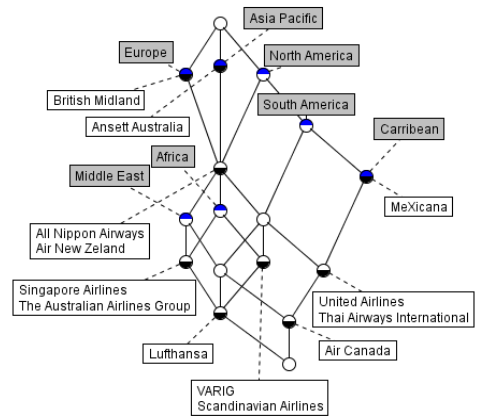

| Star Alliance | South America | Europe | Asia Pacific | Middle East | Africa | Caribbean | North America |
|---|---|---|---|---|---|---|---|
| Air Canada | × | × | × | × | | × | × |
| Air New Zealand | | | × | × | | | × |
| All Nippon Airways | | | × | × | | | × |
| Ansett Australia | | | × | | | | |
| Austrian Airl. Group | | × | × | × | × | | × |
| British Midland | | × | | | | | |
| Lufthansa | × | × | × | × | × | | × |
| Mexicana | × | | | | | × | × |
| Scandinavian Airl. | × | × | × | | × | | × |
| Singapore Airlines | | | × | × | × | × | × |
| Thai Airways Int. | × | × | × | | | × | × |
| United Airlines | × | × | × | | | × | × |
| VARIG | × | × | × | | × | | × |

**Fig. 5.** Generalizing attributes

In the following we formalize the way generalized patterns are produced. Let $\mathbb{K} := (G, M, I)$ be a context. The attributes of $\mathbb{K}$ can be grouped together to form another set of attributes, namely $S$, to get a context where the attributes are more general than in $\mathbb{K}$. For the Star Alliance example, each member is flying to an airport located in a city. Cities are generalized to countries and sometimes to regions or continents. Formally, $S$ can be seen as an index set such that $\{M_s \mid s \in S\}$ covers $M$. The context $(G, M, I)$ is then replaced with a context $(G, S, J)$ as in the scaling process. There are mainly three ways to express the binary relation $J$ between the objects of $G$ and the (generalized) attributes of $S$:

(∃) $gJs :\iff \exists m \in s, gIm$. Consider an information table describing companies and their branches in North America. We first set up a context whose objects are companies and whose attributes are the cities where these companies have or may have branches. If there are too many cities, we can decide to group them in provinces (in Canada) or states (in USA) to reduce the number of attributes. Then, the (new) set of attributes is now a set $S$ whose elements are states and provinces. It is quite natural to say that a company $g$ has a branch in a province/state $s$ if $g$ has a branch in a city $m$ which belongs to the province/state $s$. Formally, $g$ has attribute $s$ iff there is $m \in s$ such that $g$ has attribute $m$.

(∀) $gJs :\iff \forall m \in s, gIm$. Consider an information system about Ph.D. students and the components of the comprehensive doctoral exam. Assume that components are: the written part, the oral part, and the thesis proposal, and that a student succeeds in his exam if he succeeds in the three components of the exam. The objects of the context are Ph.D. students and the attributes are the different exams taken by students. If we group together the different components, for example

$$CE.written, CE.oral, CE.proposal \mapsto CE.exam,$$

then it becomes natural to say that a student $g$ succeeds in his comprehensive exam $CE.exam$ if he succeeds in *all* the exam parts of $CE$. i.e $g$ has attribute $CE.exam$ if for all $m$ in $CE.exam$, $g$ has attribute $m$.

($\alpha$%) $gJs :\iff \frac{|\{m \in s \mid gJm\}|}{|s|} \geq \alpha_s$ where $\alpha_s$ is a threshold set by the user for the generalized attribute $s$. This case generalizes the (∃)-case (take $\alpha = \frac{1}{|M|}$) and the (∀)-case (take $\alpha = 1$). To illustrate this case, let us consider a context describing different specializations in a given Master degree program. For each program there is a set of mandatory courses and a set of optional ones. Moreover, there is a predefined number of courses that a student should succeed to get a degree in a given specialization. Assume that to get a Master in Computer Science with a specialization in "computational logic" (CL), a student must have seven courses from a set $s_1$ of mandatory courses and three courses from a set $s_2$ of optional ones. Then we can introduce two generalized attributes $s_1$ and $s_2$ so that a student $g$ succeeds in the group $s_1$ if he succeeds in at least seven courses from $s_1$, and succeeds in $s_2$ if he succeeds in at least three courses from $s_2$. So, for $\alpha_{s_1} := \frac{7}{|s_1|}$ and $\alpha_{s_2} := \frac{3}{|s_2|}$, we have

$$gJs_i \iff \frac{|\{m \in s_i \mid gJm\}|}{|s_i|} \geq \alpha_{s_i}, \ 1 \leq i \leq 2.$$

Attribute generalization reduces the number of attributes. One may therefore expect a reduction of the number of concepts (i.e., $|\mathfrak{B}(G, S, J)| \leq |\mathfrak{B}(G, M, I)|$). Unfortunately, this is not always the case, as we can see from example in Fig. 6 below. Therefore, it is interesting to investigate (in the future) under which condition generalizing patterns reduces the size of the initial lattice. Moreover, finding the connections between the implications of the generalized context and the initial one is also an important open problem to be considered.

| $\mathbb{K}$ | $m_1$ | $m_2$ | $m_3$ | $m_4$ |
|---|---|---|---|---|
| $g_1$ | | | × | × |
| $g_2$ | | × | | × |
| $g_3$ | × | | × | |

| $\mathbb{K}_{\text{gen}}$ | $m_{12}$ | $m_3$ | $m_4$ |
|---|---|---|---|
| $g_1$ | | × | × |
| $g_2$ | × | | × |
| $g_3$ | × | × | |

**Fig. 6.** A generalization increasing the size of the lattice. $|\mathfrak{B}(\mathbb{K})| = 7 < 8 = |\mathfrak{B}(\mathbb{K}_{\text{gen}})|$.

A similar reasoning can be conducted with objects (rather than attributes) to replace some (or all) of them with generalized objects or clusters[10]. In such a case, the extraction can be done using an assembly of lattices having the same set of attributes, followed by a selection on generalized objects.

## 6 Approximation of Presumed Concepts: Real World concepts vs. Formal Concepts

While analyzing data, one may have a particular interest in a pair $(X, Y)$ of objects/attributes. $X$ can be for example some Star Alliance members and $Y$ some destinations. The interest of the analyst is then represented by the pair $(X, Y)$, here called *presumed concept*, as perceived by a user. A crucial question is then how a *presumed concept* can be approximated by formal concepts of a context $\mathbb{K}$. Given a presumed concept $c := (X, Y)$, we set $p\_ext(c) := X$ and $p\_int(c) := Y$ to mean the *presumed* "extent" and "intent" of $c$, respectively. We assume that there is a formal context $\mathbb{K} := (G, M, I)$ whose object set contains $p\_ext(c)$ and whose attribute set contains $p\_int(c)$. To approximate $c$, we first look at those concepts whose extent (resp. intent) has a non empty intersection with $p\_ext(c)$ (resp. $p\_int(c)$). This is done via selective (resp. projective) representations.

A presumed concept $c := (X, Y)$ has a *projective conceptual representation* $\pi_c : (A, B) \mapsto (f_2(B \cap p\_int(c)), B \cap p\_int(c))$, and a *selective conceptual representation* $\xi_c : (A, B) \mapsto (A \cap p\_ext(c), f_1(A \cap p\_ext(c)))$, defined on formal concepts $(A, B)$ of $\underline{\mathfrak{B}}(\mathbb{K})$. The inverse image of $c$ under $\pi_c$ (i.e. $\pi_c^{-1}(c)$) contains all formal concepts whose intents contain the $p\_intent$ of $c$; the inverse image of $c$ under $\xi_c$ (i.e. $\xi_c^{-1}(c)$) contains all formal concepts whose extents contain the $p\_extent$ of $c$. In fact $\pi_c^{-1}(c)$ is a principal ideal of $\underline{\mathfrak{B}}(\mathbb{K})$ generated by $H(c) := (p\_int(c)', p\_int(c)'')$, and $\xi_c^{-1}(c)$ is a principal filter of $\underline{\mathfrak{B}}(\mathbb{K})$ generated by $L(c) := (p\_ext(c)'', p\_ext(c)')$. Then $L(c)$ is the smallest concept whose extent contains the $p\_extent$ of $c$ and $H(c)$ is the largest concept whose intent contains the $p\_intent$ of $c$. Therefore $L(c)$ (respectively $H(c)$) is the extensional (resp. intensional) approximation for $c$. If $c$ is a concept, then $L(c) = c = H(c)$.

A very interesting case is when $c$ is a preconcept of $\underline{\mathfrak{B}}(\mathbb{K})$ [28]; (i.e. $X \subseteq Y'$ or equivalently $Y \subseteq X'$). Then $X \times Y \subseteq I$ and $L(c) \leq H(c)$. In this case $c$ is a rectangle full of crosses and any element of $[L(c), H(c)]$ is a maximal rectangle full of crosses that contains $c$. An approximation of $c$ is then the interval $[L(c), H(c)]$.

A presumed concept $c$ can be erroneous. In this case, $c$ is not a preconcept of $\mathbb{K}$. There are some instances in the $p\_extent$ of $c$ that do not have all attributes in the $p\_intent$ of $c$ and some attributes in the $p\_intent$ of $c$ that do not belong to all the instances of the $p\_extent$ of $c$. We call such a presumed concept *degenerated* since there is no formal concept of $\mathbb{K}$ that covers $c$.

Let assume that $c$ is a preconcept. Then, the interval of all concepts containing $c$ can be computed based on filters and ideals. We first compute the closure $X''$ of the $p\_extent$ of $c$. We get the concept $L(c) = (X'', X')$, *i.e.*, the smallest concept containing $c$. The filter $\uparrow L(c)$ contains all concepts whose extent is larger than $p\_extent(c)$. Next, we compute the closure $Y''$ of the $p\_intent$ of $c$, and get the concept $H(c) = (Y', Y'')$, that is the largest concept containing $c$. The ideal $\downarrow H(c)$ contains all concepts whose intent is larger than $p\_intent(c)$. The approximation of $c$ is then $[L(c), H(c)] = \uparrow L(c) \cap \downarrow H(c)$ and can be read easily on the lattice $\underline{\mathfrak{B}}(\mathbb{K})$.

Figure 7 shows that the preconcept $c = (\{$Air Canada, Lufthansa$\}, \{$Canada, Europe$\})$ is not a formal concept and highlights the interval $[L(c), H(c)]$ that approximate $c$ where $L(c) = (\{$Air Canada, Lufthansa$\}, \{$Canada, Europe,*Middle East, Asia-Pacific, US, Latin America, Mexico*$\})$ and $H(c) = (\{$Air Canada, Lufthansa, *Air New Zealand, The Austrian Airlines Group, Singapore Airlines, United Airlines*$\}, \{$Canada, Europe, *Asia-Pacific, US*$\})$.

## 7 Related Work

Three main approaches towards handling the overwhelming size of mined knowledge (mainly rules) are proposed: (i) constrained-based rule mining which aims to reduce the DM output by imposing constraints on the premise or the consequent of association rules [14,17], (ii) rule filtering using quality measures (e.g., support) and concise representations [23,12], and (iii) querying the DM output [4,3,2,6,11,24]. In [3], the selection and discovery of actionable formal concepts from a pattern basis are studied, using constrained-based data mining and fault-tolerant pattern generation procedures.

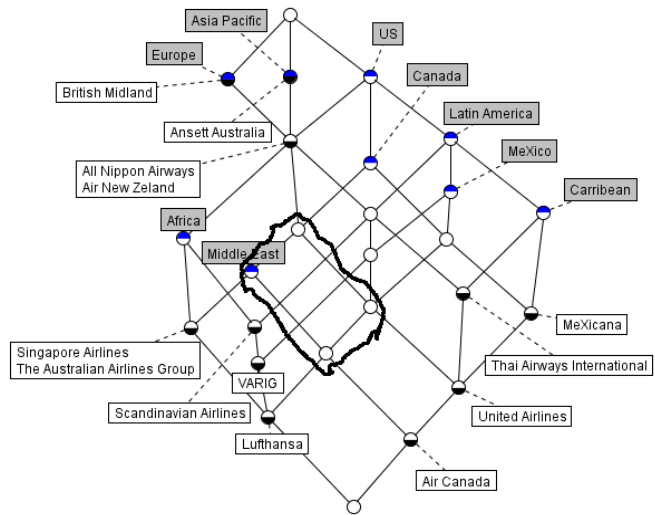

**Fig. 7.** Approximation of a preconcept in the lattice shown in Figure 2.

Recent studies on pattern management [5,24] provide a uniform framework to data and pattern management and define links between data and pattern spaces through bridging operations and cross-over queries such as finding data covered by a given pattern or identifying patterns related to a data set. Although many studies limit the management of patterns to association rules only, work conducted by Calders *et al.* [5], and Terrovitis *et al.* [24] cover different types of patterns. In [24], a pattern base management system is defined for storing, processing and querying patterns. Moreover, languages for pattern definition and manipulation are proposed, and temporal aspects of patterns are handled. In [5], a data mining algebra and a 3-World model are defined, as well as a small set of data mining primitive operators are proposed to further formulate complex queries. The proposed model includes three worlds: *D-World* for data definition and manipulation (e.g., projection, join), *I-World* for region (set of constraints) definition and manipulation, and *E-World* for operations on data contained in regions. On the industry side, work was mainly done to design languages for pattern description, manipulation and exchange (e.g., PMML).

## 8  Conclusion

In this paper we have proposed a set of operators for filtering, manipulating, and approximating a set of concepts using formal concept analysis. To date, we have analyzed and implemented four algebraic operations on lattices: the selection on a lattice according to a conjunctive condition on attributes, the projection of a lattice on a set of attributes, the assembly of two lattices related to a same set of objects, and the approximation of a presumed concept $c$ within a given related lattice.

Our future work concerns the following issues: (i) enrich the defined algebraic operators on concepts with additional ones borrowed from FCA theory, (ii) define new cross-over (mapping) operations between a data space (e.g., a relational table) and a pattern space expressed by a set of concepts, and (iii) study variants of the two kinds of mappings with their corresponding actual benefits in real-life applications.